
\magnification = 1020
\baselineskip16pt
\abovedisplayskip 3pt 
\belowdisplayskip 3pt 
\raggedbottom
\vsize=9.65 true in
\hsize=5.975 in 
\voffset= - 0.1 true in
\font\bmit=cmmib10 \textfont9=\bmit \def\bmit{\fam9 }
\font\bx=cmbsy10 \textfont10=\bx \def\bx{\fam10 }
\mathchardef\pi="7119
\mathchardef\rho="711A
\mathchardef\sigma="711B
\mathchardef\mu="7116
\mathchardef\nabla="7272
\parskip=8pt plus3pt
\interlinepenalty = 100
\def\boxit#1{\vbox{\hrule\hbox{\vrule\kern4pt
   \vbox{\kern4pt#1\kern4pt}\kern4pt\vrule}\hrule}}
\def\sqr#1#2{{\vcenter{\vbox{\hrule height.#2pt
   \hbox{\vrule width.#2pt height#1pt  \kern#1pt
      \vrule width.#2pt}
     \hrule height.#2pt }}}}


 \leftline{\bf Two particle realisation of the Poincare group
with interaction   \hfil\break  }

\rightline{Shaun N Mosley,\footnote{${}^*$} {E-mail:
shaun.mosley@ntlworld.com  }
Sunnybank, Albert Rd, Nottingham NG3 4JD, UK }

\beginsection Abstract

A relative position 4-vector is constructed for two
spin-zero particles. Some advantages of this relative position
over Bakamjian-Thomas are pointed out. The centre-of-mass (CM)
and relative positions and momenta are an explicit realisation
of the so-called non-canonical covariant representation.
The Hamiltonian including potential terms is factorised into
CM and relative components, the latter is
a Lorentz scalar readily evaluated in the CM
rest frame when the relative position, momentum are
canonical conjugates.

\beginsection Introduction

   In the non-relativistic mechanics of two particle systems,
   the conversion of the individual
particle generators into centre-of-mass (CM) and relative components
has been a fruitful concept. Instead of the
$ ( {\bf x}_i , {\bf p}_i ) \, $
and $ ( {\bf x}_j , {\bf p}_j ) \, $
conjugate pairs, one constructs the new conjugate pairs
$ ( {\bf X} , {\bf P} ) \, $ and $ ( {\bf \bar x} , {\bf p} ) \,
,$  then the Hamiltonian can be written as the sum of CM and
relative  parts. Potentials which are scalar functions of
$ {\bf \bar x} \equiv ( {\bf x}_i - \,  {\bf x}_j ) \, $ can be
inserted into the Hamiltonian such that the Galilei group
algebra is maintained, and the two body problem is effectively
reduced to one body with a potential.

In relativistic mechanics this procedure is more difficult. A
relative position 4-vector\footnote{${}^1$} {a 4-vector $
a^\lambda $ must satisfy $ \{ J^{\lambda \mu} ,a^\nu \} = \eta^{\mu
\nu} a^\lambda -  \eta^{\lambda \nu} a^\mu  \, .$}
$ q_{{}_{BT}} \, $  was found by Bakamjian and Thomas [1,2]
which will be the starting point of our discussion.
We will use covariant notation throughout, and
all the CM and relative generators will be defined in
terms of the individual particle generators
$ j_i^{\lambda \mu }, j_j^{\lambda \mu } ,  p_i^{\lambda } ,
p_j^{\lambda } \, .$
Our central task is to introduce interaction terms into the
two-particle Poincare generators
$$ \eqalign{
J^{\lambda \mu } &\equiv j_i^{\lambda \mu } + \,
j_j^{\lambda \mu } \; , \qquad
P^{\lambda } \equiv p_i^{\lambda } + \, p_j^{\lambda } \cr
   } \eqno (1.1) $$
   (where
   $ j_i^{\lambda \mu },  p_i^{\lambda } $ are the free particle
generators for particle $ i \, $ etc) such the group algebra is
maintained.  More specifically we introduce interaction terms
into the system energy-momentum 4-vector, ie $ P^\lambda
\rightarrow P_{\rm int}^\lambda \, ,$ so that the usual
relations characteristic of the Poincar\'e group still hold:
$$\eqalignno {
\{J^{\lambda \mu} , J^{\nu \rho}  \}
&= \eta^{\lambda \rho}
J^{\mu \nu} + \eta^{\mu \nu} J^{\lambda \rho} - \eta^{\lambda
\nu} J^{\mu \rho} - \eta^{\mu \rho} J^{\lambda \nu} \cr
\{ J^{\lambda \mu} , P_{\rm int}^\nu \}
&= \eta^{\mu \nu} P_{\rm int}^\lambda -
\eta^{\lambda \nu} P_{\rm int}^\mu & (1.2)  \cr
\{P_{\rm int}^\lambda ,P_{\rm int}^\mu \} &= 0 \, . & (1.3) \cr }  $$
We use the classical Poisson brackets (Pb's)
instead of commutators at this stage, to avoid the technicalities
of operator ordering. On quantisation the Pb's are
turned to commutators and a factor of $ i $ included.

We use the following notation. For two 4-vectors $ a , b
\, :$ $  \, ( a \wedge b )^{\lambda \mu } \equiv a^\lambda  b^\mu
 - \, a^\mu b^\lambda \, ;  \; a \cdot b \equiv a^\lambda
b_\lambda , \, \, \eta^{00} = 1 = - \eta^{aa} \, , \,
\eta^{\lambda \mu } = 0 $ for $ \lambda \neq \mu \, .$    \qquad

\beginsection 2 Relative position 4-vectors

The various 4-vector relative positions
$  q^\lambda \, $ to be considered in this
paper all satisfy the Poisson bracket relation
$$ \{ q^\lambda \, , \, \hat P^\mu \} = 0 \,  \eqno (2.1)  $$
 where $ \hat P^\mu \equiv P^\mu / | P | \, .$
 It then follows that we are able to insert a potential $ V \, $
 being a
function of   $  q  \, $ into the system
energy-momentum generators $ P^{\lambda } $ as follows
$$ P^{\lambda } \rightarrow  P_{\rm int}^\lambda
\equiv \hat P^{\lambda } \Big( | P | \,
   + V ( q \cdot q ) \, \Big) \eqno (2.2) $$
   and due to (2.1) the new $ P_{\rm int} ^\lambda $ interaction
``Hamiltonians''  will have zero Pb with
each other:
$$ \{ P_{\rm int}^\lambda \, , \, P_{\rm int}^\mu \}  = 0 \, ,
\eqno (2.3) $$
and
$$ \{ J^{\lambda \mu} , P_{\rm int}^\nu \}
= \eta^{\mu \nu} P_{\rm int}^\lambda -
\eta^{\lambda \nu} P_{\rm int}^\mu  $$
also follows if the potential $ V \, $ is a scalar
function of   $  q  \, .$
The Lorentz generators $ J^{\lambda \mu } $ do not contain
interaction terms, remaining the same as in (1.1).  Furthermore
we will show that the $ | P | \, $
in (2.2) can be expressed as a scalar
function of the relative momentum $ v^\lambda \, ,$ then we will
have factorised the Hamiltonian $ P_{\rm int}^0 $
into CM and relative variables.

 In using 4-vector notation we
appear to have a surplus of components. But the $  q^\lambda \,
$ and its conjugate  $  v^\lambda \, $ (to be introduced in the
next section) have the property that they are orthogonal to the
system 4-momentum $ P \equiv p_i + p_j \, ,$ ie
 $$  (q \cdot P ) =  ( v \cdot P ) = 0 \, , \eqno (2.4) $$
 so that $  q^0  , \,  v^0 \, ,$  can be regarded as
dependant variables, and are zero in the system rest frame.

{\it \noindent \underbar{The Bakamjian-Thomas relative position  }
 \hfil\break }
The Bakamjian-Thomas (BT) relative position generator is (see
(5.6) of [1])
$$ \eqalignno{
&q_{{}_{BT}}^{\lambda} \equiv  { j_{i}^{\lambda \mu}
  P_{\mu } \over  p_{i} \cdot P } \,
  -  \, { j_{j}^{\lambda \mu}
  P_{\mu } \over  p_{j} \cdot P }  \, & (2.5) \cr
  }  $$
 which is by construction orthogonal to the system momentum
 $ P^\lambda ,$ a time-like 4-vector.
 This means that $ q_{{}_{BT}}^{\lambda} $ is
 space-like or null, then
 $ ( - q_{{}_{BT}}\cdot q_{{}_{BT}} ) \, $
 is positive so can be regarded as a distance squared,
 which is also Lorentz invariant. It was shown in the original paper
 [1] that in the non-relativistic limit
 $ ( - q_{{}_{BT}}\cdot q_{{}_{BT}} ) \,
 \simeq \, | {\bf x}_i - \, {\bf x}_j |^2 \, .$
 
 Let us assume that a potential function of this distance,
 $ V ( - q_{{}_{BT}}
 \cdot q_{{}_{BT}} ) \, ,$ is included in the Hamiltonian, then the
 force on particle $ i \, $ is worked out from
 $ \{ V ( - q_{{}_{BT}}\cdot q_{{}_{BT}} ) \, , \, p_i \} \, .$
 We can calculate from (2.5) and the individual particle Pb's
corresponding to (1.2,3) that
 $$ \eqalignno{
 \{ q_{{}_{BT}}^{\lambda}  \, , \, p_i^{\mu} \}
&= - \eta^{\lambda \mu}
+ \, {  p_i^{\lambda} P^\mu \over p_{i} \cdot P } \, & (2.6) \cr
 {1 \over 2 } \, \{ - q_{{}_{BT}}
 \cdot q_{{}_{BT}}  \, , \, p_i^{\mu} \}
&= \,  q_{{}_{BT}}^{\mu }
- \, { (  p_i \cdot  q_{{}_{BT}} ) P^\mu \over p_{i} \cdot P } \,
& (2.7)  \cr
&\equiv { \,  ( q_{{}_{BT}} \wedge P   )^{\mu \rho } \over
 p_{i} \cdot P } \, p_{i\rho } \, \equiv \,
 F^{\mu \rho } p_{i\rho } \, . & (2.8)  \cr
  } $$
  In the system rest frame when $ {\bf P} = 0 \, $
        the force tensor $ F^{\mu \rho } $ only has
        ``electric'' components $ ( F^{10} , F^{20} , F^{30} ) $
        and the
        force acts along the space component of $ q_{{}_{BT}} .$
    [To include ``magnetic'' forces due to the motion of particle
    $ j \, $ the numerator of $ F \, $ should be of the form
    $ ( q \wedge p_j   )^{\mu \rho } $ instead of
    $( q_{{}_{BT}} \wedge P   )^{\mu \rho } ,$ which property
    we will show for the new relative position introduced below.]

    We now introduce the 4-vector relative momentum $ v^\lambda
\, $  which is the usual relative momentum in the
non-relativistic limit:
    $$ \eqalignno{
 v^\lambda &\equiv { p_j \cdot P \over P^2 } \, p_i^\lambda
 - \, { p_j \cdot P \over P^2 } \, p_i^\lambda & (2.9) \cr
 &\qquad \simeq ( 0 \, , { m_j {\bf p}_i \, - m_i {\bf p}_i )
  \over m_i + m_j } \qquad \hbox{in the NR limit, } \cr
    \noalign{ \noindent Alternatively $ v \, $ can be written as  }
v^\lambda
&= p_{-\bot}^{\lambda} \equiv  p_{-}^{\lambda}   \,
- \, { \hat P^{\lambda}  (  p_-  \cdot \hat P )  } & (2.10) \cr
 } $$
with
$$ p_- \equiv {1 / 2 } ( p_i - p_j ) \,  .$$
 It can be shown that (see the appendix)
 $$ \eqalignno{
&\{ q_{{}_{BT}}^\lambda \, , \, v^\mu \}
= \, - \, \eta^{\lambda \mu } + {\hat P}^\lambda {\hat P}^\mu  \,
 & (2.11) \cr
  } $$
  and any such variables being orthogonal to $ P \, $
and which satisfy the relation (2.11) we will call
``covariant conjugates'', as in the system rest frame the
components $ q_{{}_{BT}}^0 , \, v^0 $ are zero, and    $
\{ q_{{}_{BT}}^a \, , \, v^b \} = \, \delta^{ab } \, $ in the
usual manner $ \, (a,b = 1,2,3).$

{\it \noindent \underbar{An alternative relative position  }
 \hfil\break }
 The construction of the  $ {q} \, $  proceeds as follows.
First we introduce the 4-vector
$$ \eqalignno{
&q^{\lambda} \equiv  { j_{i}^{\lambda \mu}
  p_{j\mu } -  \,  j_{j}^{\lambda \mu}
 p_{i\mu } \over  p_{i} \cdot p_j } \, . & (2.12) \cr
  }  $$
  However this $ q \, $ is not necessarily space-like, so
 next we project $ q \, $ onto the $ P \, $ hyperplane, defining
$$ q_\bot^\lambda \equiv  q^{\lambda} \,
- \, {\hat P}^\lambda \, ( q \cdot \hat P \big) \, . \eqno (2.13) $$
As the $ q_\bot \, $ is by construction orthogonal to $ P \, ,$
it is space-like or null and
 $ \sqrt{ - {q}_\bot^{\lambda} {q}_{\bot \lambda} }
\, $ defines a Lorentz invariant distance (which can also
be shown to be
the usual distance
$ | {\bf x}_i - \, {\bf x}_j | \, $ in the non-relativistic limit).
 We can now calculate the
following Pb identities (as shown in the appendix)
$$ \eqalignno{
 \{ {q}_\bot^{\lambda}  \, , \, p_i^{\mu} \}
&= - \eta^{\lambda \mu}
+ \,  \hat P^{\lambda}  \hat P^\mu \,
+ \, p_j^\mu \, \big( { v^\lambda  \over  p_{i} \cdot p_j } \big) \,
& (2.14)   \cr
\{ {q}_\bot^{\lambda}  \, , \, p_j^{\mu} \}
&= \, \eta^{\lambda \mu}
- \,  \hat P^{\lambda}  \hat P^\mu  \,
+ \, p_i^\mu \, \big( { v^\lambda  \over  p_{i} \cdot p_j }
\big) \, . & (2.15)  \cr
    } $$
    As above we calculate $ \{ - q_\bot
 \cdot q_\bot  \, , \, p_i^{\mu} \} \, $ for the
 force on particle $ i \, $ due to a potential
 $ V ( - q_\bot^2 ) \, :$
 $$ \eqalignno{
 {1 \over 2 } \, \{ - q_\bot \cdot q_\bot  \, , \, p_i^{\mu} \}
&= \,  q_\bot^{\mu }
- \, { (  v \cdot  q_\bot ) p_j^\mu \over p_{i} \cdot p_j } \, \cr
&= \,  q_\bot^{\mu }
- \, { (  p_i \cdot  q_\bot ) p_j^\mu \over p_{i} \cdot p_j } \,
  \cr
&\equiv { \,  ( q_\bot \wedge p_j   )^{\mu \rho } \over
 p_{i} \cdot p_j } \, p_{i\rho } \, \equiv \,
 F'^{\mu \rho } p_{i\rho } \, & (2.16)  \cr
  } $$
        and we see that the force tensor $ F'^{\mu \rho } $ now has
        ``magnetic'' components $ ( F^{23} , F^{31} , F^{12} ) $
        due to the motion of particle $ j \, ,$ and in fact
        $ F' $ is of remarkably similar form to the electromagnetic
        field produced by particle $ j \, $ if we put
        $ V \propto ( - q_\bot^2 )^{-1/2} \, $
       (see for example (14.15) of Jackson [3] ).

 Adding (2.9), (2.10) yields the Pb
relation between $ {q}_\bot , \, P  \, :$
$$ \eqalignno{
\{ {q}_\bot^{\lambda}  \, , \, P^{\mu} \}
&=  \, P^{\mu}  {  v^\lambda   \over  p_{i} \cdot p_j }
\qquad \Rightarrow \qquad \,
\{ {q}_\bot^{\lambda}  \, , \, | P | \}
=  \, | P |  {  v^\lambda   \over  p_{i} \cdot p_j }  \, &
(2.17a,b) \cr \noalign{ \noindent which means that  }   \{
{q}_\bot^{\lambda} \, , \, \hat P^{\mu} \} &\equiv \{
{q}_\bot^{\lambda}  \, , \, P^{\mu} / | P | \} = 0 \,   & (2.18)
 \cr
 }  $$
 which last relation allows us to introduce potentials into the
 Hamiltonian as discussed in Sec. 1.
 To find the Pb relation between $ {q}_\bot \, $ and
 $ v \, ,$ we first subtract (2.15) from (2.14) yielding
 $$ \eqalignno{
 \{ {q}_\bot^{\lambda}  \, , \, p_-^{\mu} \}
 &= - \eta^{\lambda \mu} \,
 + \, { \hat P^{\lambda} \hat P^\mu  } \, -  \, v^\lambda
 \big( {  p_-^{\mu} \over  p_{i} \cdot p_j } \big) \, \cr
 \{ {q}_\bot^{\lambda}  \, , \, ( p_- \cdot \hat P )  \}
 &= \, -  \, v^\lambda
 \big( { ( p_-\cdot \hat P ) \over  p_{i} \cdot p_j } \big) \, \cr
 \{ {q}_\bot^{\lambda}  \, , \, v^\mu  \}
 &\equiv \{ {q}_\bot^{\lambda}  \, , \, p_-^{\lambda}   \,
- \, { \hat P^{\lambda}  (  p_-  \cdot \hat P )  }  \}  \cr
 &= - \eta^{\lambda \mu} \,
 + \, { \hat P^{\lambda} \hat P^\mu  } \, -  \, v^\lambda
 \big( {  p_-^{\mu} \over  p_{i} \cdot p_j } \big) \,
 +  \, \hat P^{\mu} \,
 v^\lambda
 \big( { ( p_-\cdot \hat P ) \over  p_{i} \cdot p_j } \big) \, \cr
&= - \eta^{\lambda \mu} \,  + \, { \hat P^{\lambda} \hat P^\mu
} \,  - \,  { v^\lambda v^\mu \over  p_{i} \cdot p_j }  \,  .
& (2.19) \cr
   }  $$
We see that $ {q}_\bot , \, v \, ,$ do not qualify
as covariant conjugates satisfying (2.11), because of the extra
term on the RHS of (2.19).

There are two ways at arriving at a covariant conjugate pair.
The first is to define
$$ q'^\lambda \equiv {q}_\bot^{\lambda}
- \, \big( { {q}_\bot \cdot v \over v^2 + \, p_i \cdot p_j } \big)
\, v^\lambda \eqno (2.20)  $$
  then using (2.19) it can be readily shown that
$$ \eqalignno{
 \{ {q}'^{\lambda}  \, , \, v^\mu  \}
 &= - \eta^{\lambda \mu} \,
 + \, { \hat P^{\lambda} \hat P^\mu  } \,    &(2.21)   \cr
   }  $$
   so that $ q' , v \, $ are covariant conjugates satisfying
(2.11).  But it turns out that $ q' $ is none other that the
Bakamjian-Thomas relative position that we have already
encountered: the fact that
 $$ q' \equiv q_{{}_{BT}} $$
 which is rather a tedious calculation is shown in the appendix.

 The second way to arrive at a covariant conjugate pair - which
we will follow for the rest of this paper - is to rescale  $
q_\bot , v \, $ and define
  $$\eqalignno{
 &  \rho  \equiv  { {q}_\bot \over | P | } \, ,
  \qquad   \pi  \equiv   | P | v & (2.22)  \cr
  \noalign{ \noindent then  inspection of (2.12b) tells us that
   $ \rho , \, \pi \, ,$ have the
required Pb relation  }
&\{  \rho^\lambda \, , \, \pi^\mu  \}
= - \eta^{\lambda \mu} \,
 + \, { \hat P^{\lambda} \hat P^\mu  } \, .
 & (2.23)   \cr
   }  $$
   One reason for preferring the pair $ ( \rho , \pi ) \, $
   for the relative or internal variables is that potential
   functions of $ \rho \, $ have the attractive feature of
   producing electromagnetic type forces (recalling (2.16)),
instead of the pure electric type forces resulting from
the BT relative position. In the next Section we will
explore additional reasons for adopting  $ ( \rho , \pi ) \, $
instead of $ ( q_{{}_{BT}}, v ) \, $  as the relative position
and relative momentum 4-vectors, including \hfil\break
(1) it is easier
  to factorise the Hamiltonian into CM and relative components
when the relative component is expressed in terms of    $ \pi \,
$ rather than $ v \, .$ And \hfil\break
(2) the CM and relative variables when these latter are expressed
in terms of  $ ( \rho , \pi ) \, $ are maximally independent and
are an explicit realisation of the so-called `non-canonical
covariant realisation' (NCR) of [4], as discussed in Sec 4.

\beginsection 3. Factorisation of the  Hamiltonian
$ P_{\rm int }^0 $

  As discussed at the beginning of Sec 2, the relation (2.1)
allows us to introduce an interaction potential $ V \, $ being a
scalar function of $ - \rho^2 $ into the Hamiltonian
$$ P_{\rm int }^0
= \hat P^0 \Big( | P | + V (- \rho^2 ) \Big) \, \eqno (3.1) $$
while
maintaining the Pb relations of the Poincare group algebra. In
this section we will first show that the  $ | P | \, $ component
in the Hamiltonian can be written in terms of $ \pi^2 ,$ thus
achieving a factorisation of the $ P_{\rm int }^\lambda $ into
CM and relative generators. From (2.22), (2.10)
$$ \eqalignno{
  \pi^2
 &\equiv P^2 \, v^2 = P^2 [ p_-^2 - \, ( \hat P \cdot p_- )^2 ]  \,
\cr
&=  {1 \over 4 } \, \big[ ( m_i^2 + m_j^2 + 2 p_i \cdot p_j )
 ( m_i^2 + m_j^2 - 2 p_i \cdot p_j ) \, - \, ( m_i^2 - m_j^2 )^2 \big]
 \, \cr
 &=  m_i^2  m_j^2 - \, ( p_i \cdot p_j )^2 \, \cr
  \noalign{ \noindent or }
   p_i \cdot p_j &= [ m_i^2 m_j^2 - \, \pi^2 ]^{1/2} & (3.2)
\cr
 \noalign{ \noindent recalling that $ - \pi^2  $ is positive
 due to $ \pi \cdot P = 0 \, .$
 Then }
 | P | &= \big[  m_i^2 + m_j^2 + 2 p_i \cdot p_j  \big]^{1/2} \cr
 &= \big[  m_i^2 + m_j^2 + 2 [ m_i^2 m_j^2 - \, \pi^2 ]^{1/2}
  \big]^{1/2}\, .  & (3.3) \cr
  }  $$
  In the non-relativistic regime when $ - \pi^2
  \ll m_i^2 m_j^2 ,$ then
  $$ \eqalignno{
  | P |
  &\simeq \big[  m_i^2 + m_j^2 + 2 m_i m_j
   - \, { \pi^2 \over m_i m_j }  \big]^{1/2} \cr
  &\simeq  (  m_i + m_j )  - \, {1 \over 2} { \pi^2
   \over  2 m_i m_j  (  m_i + m_j ) }  \cr
   &\qquad =  (  m_i + m_j )  - \, {1 \over 2} {1 \over \mu}
   { \pi  \over  (  m_i + m_j ) } \cdot
   { \pi  \over  (  m_i + m_j ) } &            (3.4)\cr
    }  $$
    where $ \mu \, $ is the usual reduced mass $ \mu
    = { m_i m_j \over m_i + m_j } \, .$  This is the standard
non-relativistic expression for the energy taking into account that
from (2.22)  $  \pi \, $ has been rescaled by a factor of
$ | P |  \simeq m_i + m_j \, ,$
and that in the CM rest frame  $ -  \pi^\lambda \pi_\lambda
\rightarrow \, +  \pi^a \pi^a \, = {\bmit \pi}^2 \; (a = 1,2,3).$

 Recalling the interaction Hamiltonian (3.1),
 the relative (internal) factor
 $$  \Big( | P | + V ( \rho ) \, \Big) \,  $$
 is a Lorentz scalar having the same
value in any frame. This means that we can
choose to evaluate it in the CM
frame, when the $ {\bmit \rho } , \, {\bmit \pi } \, $ are
conjugate variables, and we can follow the usual procedures in
quantising the internal component of the Hamiltonian as discussed
in Sec 5.

The CM factor $ \hat P^0 $ in (3.1) is just the usual
relativistic dilation factor $ \gamma \, ,$ which means that the two
interacting particles are indeed behaving as one system. Bakamjian
[5] noted the advantages of
including the interaction terms in the 4-momentum vector (rather
than in the boost generator as in the original BT paper) on
physical grounds, in that the energy of interaction effectively
increases the system rest mass, which in turn must contribute to
the system momentum.

 \beginsection 4 Relations between the CM and relative variables
 
 The CM generators are $ R , \hat P \, $ where
 $$ R^\lambda \equiv J^{\lambda \rho } \hat P_\rho \, ,
 \qquad \hat P^\lambda \equiv  P^\lambda / |P| \, \eqno (4.1) $$
 recalling from (1.1) that $ J \equiv j_i + j_j \, , \;
 P \equiv p_i + p_j \, .$ The $ R^\lambda $ is essentially the
Shirokov  position 4-vector [6] (but
note that the Shirokov 4-vector position is
$ J^{\lambda \rho } \hat P_\rho / P^2 \, ,$ we have multiplied it by
 $ | P | \, $ to obtain $ R \, ,$ so that
$ R \, $ is the covariant
conjugate to $ \hat P \, $ rather than $ P \, ).$

 The relative generators are
 $ \rho , \, \pi \, $ which were defined in (2.22) and (2.13).
 The CM and relative generators are not in general independent as
in the non-relativistic case, except in the system rest frame.
The Pb relations below follow from the definitions of the
generators in terms of $ j_i , j_j , p_i , p_j ,$ as an example
we calculate $ \{ R^\lambda \, , \, \rho^\mu \} \, $
in the appendix.
 $$\openup1\jot \tabskip = 0 pt plus1fil
\halign to\displaywidth{
$\displaystyle#$&\quad${}\displaystyle#\hfil$&
\qquad$\displaystyle#\hfil$ \tabskip = 0 pt plus1fil&
\llap {#}\tabskip = 0 pt\cr
&\{ R^\lambda \, , \, \hat P^\mu \} \, = \, - \, \eta_{\lambda \mu }
+ \, \hat P^\lambda \, \hat P^\mu
\; &\{ \rho^\lambda  \, , \, \pi^\mu \} \,
= \, - \, \eta_{\lambda \mu } + \, \hat P^\lambda \, \hat P^\mu
\;  &  (4.2)  \cr
&\{ R^\lambda \, , \, R^\mu \} \,
= \, J^{\lambda \mu }
\; &\{ \rho^\lambda  \, , \, \rho^\mu \}
\, = \, 0 \, \;  &  (4.3)  \cr
&\{ \hat P^\lambda \, , \, \hat P^\mu \} \,
= \, 0 \;
&\{ \pi^\lambda \, , \, \pi^\mu \} \, = \, 0 \;
 &  (4.4)  \cr
 \noalign{ also the cross terms }
 &\{ R^\lambda \, , \, \pi^\mu \} \, = \,  \hat P^\mu \,
\pi^\lambda  \;
&\{ \hat P^\lambda  \, , \, \rho^\mu \}
\, = \, 0 \;  &  (4.5)  \cr
&\{ R^\lambda \, , \, \rho^\mu  \} \,
= \, \hat P^\mu \, \rho^\lambda \;
&\{ \hat P^\lambda  \, , \, \pi^\mu  \} \, = \, 0 \, .  &  (4.6)
\cr
   } $$
   In the system rest frame when $ {\bf P} = 0 \, $ then
   $ R^0 = \rho^0 = \pi^0 = 0 \, ,$ and the only Pb's above which
are non-zero are
$$ \{ R^a \, , \, \hat P^b \} \, = \, \delta^{a b }  \, , \quad
\{ \rho^a  \, , \, \pi^b \} \,
= \, \, \delta^{a b }  \, , \quad \{ R^a \, , \, R^b \} \,
= \, J^{a b }  \qquad \qquad \hbox{when $ {\bf P} = 0 \, .$ } $$
so that in this case $  {\bmit \rho } , \,  {\bmit \pi }  \, $
are conjugates to each other in the usual sense.

We can split the Lorentz generators $ J^{\lambda \mu } $
into external and relative parts $ L^{\lambda \mu } , \,
 M^{\lambda \mu } $  such that
  $$\eqalignno{
L^{\lambda \mu} \, + \, M^{\lambda \mu}
 &= \, J^{\lambda \mu}  & (4.7) \cr
   }  $$
   where $ L^{\lambda \mu } , \,  M^{\lambda \mu } $ are
 $$\eqalign{
L^{\lambda \mu} &= \, (  R \wedge \hat P )^{\lambda \mu } \, \cr
M^{\lambda \mu} &= \, (  \rho \wedge \pi )^{\lambda \mu } \, .
\cr
  } \eqno (4.8)  $$
  Note that (4.7) only holds for zero-spin particles, additional
terms are required to make up the total angular momentum $ J \, $
if one or more of the particles has spin.
For the Pb relations involving $ L , M , $ we introduce the
shorthand
$$ \eta_\bot^{\lambda \mu } \equiv  \eta^{\lambda \mu }
- \, \hat P^\lambda  \hat P^\mu        \eqno (4.9) $$
Then from the above relations (4.2-6) it follows that
$$ \eqalignno{
&\{ L^{\lambda \mu } \, , \, R^\rho \} \,
= \, \eta^{\mu \rho }  R^\lambda
- \, \eta^{\lambda \rho }  R^\mu \,
- \, M^{\mu \rho } \hat P^\lambda \,
+ \, M^{\lambda \rho } \hat P^\mu  \cr
& \{ M^{\lambda \mu } \, , \, R^\rho \} \, =  M^{\mu \rho }
\hat P^\lambda \,
- \, M^{\lambda \rho } \hat P^\mu \, \;  &  (4.10)  \cr
 &\{ L^{\lambda \mu } \, , \, \hat P^\rho \} \, = \, \eta^{\mu
\rho }  \hat P^\lambda - \, \eta^{\lambda \rho }  \hat P^\mu \, \cr
&\{
M^{\lambda \mu } \, , \, \hat P^\rho \} \, = \, 0 \;  &  (4.11)  \cr
&\{ L^{\lambda \mu } \, , \, \rho^\rho \} \,  = \,
\hat P^\rho \, \big( \rho \wedge \hat P \, \big)^{\lambda
\mu } \; \cr
&\{ M^{\lambda \mu } \, , \, \rho^\rho \}
\, = \,  \eta_\bot^{\mu \rho }  \,
\rho^\lambda \, - \,  \eta_\bot^{\lambda \rho } \,
\rho^\mu  \;  &  (4.12)  \cr
&\{ L^{\lambda \mu } \,
, \,  \pi^\rho \} \,  = \, \hat P^\rho \, \big( \pi \wedge \hat
P \, \big)^{\lambda \mu } \; \cr
&\{ M^{\lambda \mu } \, , \, \pi^\rho \} \,
= \,  \eta_\bot^{\mu \rho }  \, \pi^\rho \,
 - \, \eta_\bot^{\lambda \rho } \, \pi^\mu  \;  &  (4.13) \cr
   } $$
 Adding the $ L^{\lambda \mu } , \,
 M^{\lambda \mu } $ pairs above, all generators have the correct
 4-vector Pb relations with $ J^{\lambda \mu } \, .$
 
     Finally from the above we can readily determine the Pb
relations $ \{L^{\lambda \mu} \, , \, L^{\nu \rho}  \} , \,
\{L^{\lambda \mu} \, , \, M^{\nu \rho}  \} \, $ and $ \,
\{M^{\lambda \mu} \, , \, M^{\nu \rho}  \} \, .$ In
particular
$$\eqalignno{
\{M^{\lambda \mu} \, , \, M^{\nu \rho}  \}
&=  \eta_\bot^{\lambda \rho}  M^{\mu \nu}
+ \, \eta_\bot^{\mu \nu} M^{\lambda \rho}
- \, \eta_\bot^{\lambda \nu}  M^{\mu \rho}
-  \eta_\bot^{\mu \rho}  M^{\lambda \nu}\, . & (4.14) \cr
  }  $$
Note that $ M \equiv ( \rho \wedge \pi ) \, $ is orthogonal to $
P \, ,$ i.e  $ M^{\lambda \mu} P_\mu = 0 \, ,$ which is necessary
for the system to have
space inversion invariance [4] and is a consequence of
    both $ \rho , \, \pi \, $ being orthogonal to $ P \, .$  In
the CM rest frame only the components
$$ \{ M^{23}, M^{31},M^{12} \} \rightarrow \{ S^1, S^2, S^3 \} \, $$
exist, and from (4.14) the
$ {\bf S } \, $ obey the same Pb's as the usual
spin 3-vector. Also from (4.11) we see that
$ M^{\lambda \mu} $ is translation invariant as required for it
to represent an internal angular momentum.

The Pb relations relations above mean that $ R , \hat P , \rho ,
\pi \, $ are an explicit realisation of the algebra which
Rohrlich [4] labelled the non-canonical covariant realisation
(NCR). We have constructed $ R , \hat P , \rho ,
\pi \, $ satisfying the NCR in terms of the individual particle
generators  $ j_i , j_j , p_i , p_j .$ This realisation of the
NCR is unique.

 \beginsection 5 Quantisation and outlook
 
 Much effort was spent by previous workers, for example [7], in
finding CM and relative generators which are canonical
(meaning that $
Q , P , q , p \, $ satisfy the canonical relations $
\{ Q^a , P^b \}  =  \{ q^a , p^b \} = \delta^{ab } \, )$ in any
inertial frame. Their approach was to find the relative position
and momentum generators in the CM rest frame, then Lorentz boost
these 3-vector generators to a general frame. This results in
complicated expressions.
 
 Our approach depends on being able to factorise the Hamiltonian
into CM and relative components as in (3.1), i.e.
 $$ P_{\rm int}^0  = \hat P^0 \big( | P | + V (\rho ) \Big)
 = \hat P^0 \, {\cal H} (\rho , \pi )  \eqno (5.1) $$
where the
relative component $ {\cal H} (\rho , \pi )  $ is a Lorentz
scalar of $ \rho , \pi \, .$  We then quantise $ {\cal H} \, $
in the CM rest frame, when the $ \rho , \pi \, $  are
conjugate variables satisfying $ \{ \rho^a , \pi^b \}  =
\delta^{ab } \, ,$ and because $ {\cal H} \, $ is a Lorentz
scalar it will have the same eigenvalues in any other frame. The
CM component $ \hat P^0 \equiv P^0 / | P | $ is just the usual
relativistic dilation factor $ \gamma \, .$

Below we outline the quantisation of a two particle system with a
Coulomb potential
$$ V = \, - \, ( - q_\bot \cdot q_\bot )^{-1/2} = \, - \,
( - \rho \cdot \rho )^{-1/2} |P|^{-1} \, $$
recalling from (2.22) that $ q_\bot = |P| \rho \, .$
Then
$$ \eqalignno{
 {\cal H} &\equiv | P | + V  = | P | \, - \,
  ( - \rho \cdot \rho )^{-1/2} \, |P|^{-1}  \cr
 &= \big[  m_i^2 + m_j^2 + 2 [ m_i^2 m_j^2 - \, \pi^2 ]^{1/2}
  \big]^{1/2} \,  \, - \,  ( - \rho \cdot \rho )^{-1/2}
  \big[  m_i^2 + m_j^2 + 2 [ m_i^2 m_j^2 - \, \pi^2 ]^{1/2}
  \big]^{-1/2}\,   \cr
  \noalign{ \noindent and in the CM frame
  $ \; - \rho \cdot \rho
  \rightarrow {\bmit \rho}^2 , \;  - \pi \cdot \pi
  \rightarrow {\bmit \pi}^2 ,$ then }
  {\cal H} &= \big[  m_i^2 + m_j^2 + 2 [ m_i^2 m_j^2
  + \, {\bmit \pi}^2 ]^{1/2}
 \big]^{1/2} \, - \,  |{\bmit \rho}|^{-1}
\big[  m_i^2 + m_j^2 + 2 [ m_i^2 m_j^2 + \, {\bmit \pi}^2 ]^{1/2}
  \big]^{-1/2}\,  & (5.2)  \cr
  }  $$
 where $ \rho , \pi \, $  are
conjugate variables satisfying $ \{ \rho^a , \pi^b \}  =
\delta^{ab } \, .$ Putting $ {\cal H} = E \, ,$ this eigenvalue
equation can be solved numerically if not in closed form. Further
aspects of quantisation will be addressed elsewhere.

\beginsection Appendix

{\it \noindent \underbar{Proof of (2.11)  }
 \hfil\break }
 First we calculate
 $ \{ q_{{}_{BT}}^{\lambda}  \, , \, p_i^{\mu} \} \, :$
  $$ \eqalignno{
 \{ q_{{}_{BT}}^{\lambda}  \, , \, p_i^{\mu} \}
&\equiv  \{ { j_{i}^{\lambda \rho}
  P_{\rho } \over  p_{i} \cdot P } \,
  -  \, { j_{j}^{\lambda \rho}
  P_{\rho } \over  p_{j} \cdot P }  \, \, , \, p_i^{\mu} \}  \cr
&=  \, {  P_\rho \over p_{i} \cdot P } \,
( \eta^{\rho \mu} p_i^\lambda - \eta^{\lambda \mu} p_i^\rho )
 \cr
&= - \eta^{\lambda \mu} + \, {  p_i^{\lambda} P^\mu
\over p_{i} \cdot P } \, & \hbox{which is } (2.6) \cr
\noalign{\noindent then }
\{ q_{{}_{BT}}^{\lambda}  \, , \, {1 \over 2 } ( p_i^{\mu}
- p_j^\mu ) \}
 &= - \eta^{\lambda \mu} + \, {1 \over 2 } ( {  p_i^{\lambda}
\over p_{i} \cdot P } \,
- {  p_j^{\lambda} \over p_{j} \cdot P }  ) P^\mu  \cr
\{ q_{{}_{BT}}^{\lambda}  \, , \, v^\mu  \}
&\equiv \{ q_{{}_{BT}}^{\lambda}  \, , \, {1 \over 2 } (
p_i^{\mu}  - p_j^\mu ) \, - \, {1 \over 2 }  \hat P^\mu
[( p_i  - p_j ) \cdot \hat P ] \, \}  \cr
  &= - \eta^{\lambda \mu} + \, \hat P^\lambda \hat P^\mu
  & \hbox{which is } (2.11)  \cr
   } $$
 
 {\it \noindent \underbar{Proof of (2.14,15)  }
 \hfil\break }
  First we calculate
 $ \{ q^{\lambda}  \, , \, p_i^{\mu} \} \, $ recalling $ q \, $
 from (2.12):
  $$ \eqalignno{
 \{ q^{\lambda}  \, , \, p_i^{\mu} \}
&\equiv  \{{ j_{i}^{\lambda \mu}
  p_{j\mu } -  \,  j_{j}^{\lambda \mu}
 p_{i\mu } \over  p_{i} \cdot p_j } \,\, , \, p_i^{\mu} \}  \cr
&=  \, {  p_{j \rho } \over p_{i} \cdot p_j } \,
( \eta^{\rho \mu} p_i^\lambda - \eta^{\lambda \mu} p_i^\rho )
 \cr
&= - \eta^{\lambda \mu} + \, {  p_i^{\lambda} p_j^\mu
\over p_{i} \cdot p_j } \,  \cr
\noalign{\noindent and  }
\{ q \cdot \hat P  \, , \, p_i^{\mu} \}
&= - \hat P^\mu + \, { ( p_i \cdot P ) p_j^\mu
\over p_{i} \cdot p_j } \,  \cr
\{ q_\bot^{\lambda}  \, , \,  p_i^{\mu}  \}
 &\equiv \{ q^{\lambda}  \, - \, (  q \cdot \hat P ) \,
 \hat P^\lambda \, , \, p_i^{\mu} \} \cr
 &=  - \eta^{\lambda \mu} + \, {  p_i^{\lambda} p_j^\mu
\over p_{i} \cdot p_j } \, - \, \hat P^\lambda
[ - \hat P^\mu + \, { ( p_i \cdot P ) p_j^\mu
\over p_{i} \cdot p_j } \, ]   \cr
&=  - \eta^{\lambda \mu} + \, \hat P^\lambda \hat P^\mu
+ {   p_j^\mu \over p_{i} \cdot p_j } \,
 ( p_i^{\lambda} - \, ( p_i \cdot P ) \hat P^\lambda )   \cr
 &=  - \eta^{\lambda \mu} + \, \hat P^\lambda \hat P^\mu
+ {   p_j^\mu \over p_{i} \cdot p_j } \,
 v^{\lambda} & \hbox{which is } (2.14) \cr
   } $$

   \vskip 1 in
 
 {\it \noindent \underbar{The equivalence of $ q' \, $
 and $ q_{{}_{BT}} $ }
 \hfil\break }
 We must show that, recalling (2.20),
 $$ {q}_\bot^{\lambda}
- \, \big( { {q}_\bot \cdot v \over v^2 + \, p_i \cdot p_j } \big)
\, v^\lambda = q_{{}_{BT}}^{\lambda} \eqno (2.20)  $$
or equivalently
$$ q_{{}_{BT}}^{\lambda} + { q_{{}_{BT}} \cdot v
\over p_{i} \cdot p_j } v =  q_\bot \, . \eqno (A1) $$
The identity (A1) is most easily shown if we employ the auxillary
variable (the Shirokov position for particle $ i \, )$
$$ q_i^\lambda \equiv j_i^{\lambda \rho } p_{i \rho } / m_i^2
\qquad \Rightarrow \qquad q_i \cdot p_i = 0 \, . $$
 then
 $$ q_{{}_{BT}} = q_i - q_j - p_i
 ({ q_i \cdot P \over p_i \cdot P })
 + p_j  ({ q_j \cdot P \over p_j \cdot P }) $$
 and using identities such as
 $$ p_i = \hat P ( p_i \cdot \hat P ) + v  \; , \qquad
 p_j = \hat P ( p_j \cdot \hat P ) - v \; , \qquad
 v^2 + p_i \cdot p_j = ( p_i \cdot \hat P ) \, ( p_j \cdot \hat P )
 $$
 we finally arrive at
 $$ q_{{}_{BT}} + ({ q_{{}_{BT}} \cdot v
\over p_{i} \cdot p_j }) \,  v
=  ( q_i - q_j ) \,
- \hat P [( q_i - q_j )\cdot \hat P ]
- v \, [ {( q_i + q_j )\cdot P \over p_i \cdot p_j } ] \eqno (A2) $$
 and the RHS of (A2) can be shown to be equal to $ q_\bot $
 by expanding out similarly as above.

 {\it \noindent \underbar{To calculate
 $ \{ R^{\lambda}  \, , \, \rho^{\mu} \} \, $  }
 \hfil\break }
 Recall $ R^\lambda \equiv J^{\lambda \rho } \hat P_\rho \, ,$
 then
 $$   \eqalignno{
 \{ R^{\lambda}  \, , \, \rho^{\mu} \} \,
 &\equiv  \{ J^{\lambda \rho } \hat P_\rho \, , \, \rho^{\mu} \} \,
  \cr
 &= ( \eta^{\rho \mu} \rho^\lambda - \eta^{\lambda \mu} \rho^\rho )   \,
  \, \hat P_\rho \cr
   &= \hat P^\mu \rho^\lambda   \, & \hbox{as in } (4.6) \cr
  }  $$

    \vskip 0.25 in
\noindent  {\bf References } \hfil\break
\nobreak
\frenchspacing
\noindent
[ 1 ]  L. H. Thomas 1952
{\it The relativistic dynamics of a system of particles
interacting at a distance } Phys. Rev. {\bf 85 }, 868 \hfil\break
[ 2 ]  B. Bakamjian and L. H. Thomas 1953
{\it Relativistic particle dynamics II }
Phys. Rev. {\bf 92 }, 1300 \hfil\break
[ 3 ]  J. D. Jackson 1975 {\it Classical Electrodynamics} 2nd
edition (New York: Wiley) \hfil\break
[ 4 ]  F. Rohrlich 1979 {\it Relativistic
Hamiltonian dynamics I. Classical mechanics }
Ann. Phys {\bf 117}, 292 \hfil\break
[ 5 ]  B. Bakamjian 1961
{\it Relativistic particle dynamics } Phys. Rev. {\bf 121 },
1849 \hfil\break
[ 6 ]  M. Lorente and P. Roman  1974
{\it General expressions for the position and spin operators in
relativistic systems } J. Math. Phys. {\bf 15 }, 70 \hfil\break
[ 8 ]  M. Pauri and G. M. Prosperi 1976
{\it Canonical representations of the Poincare group II }
J. Math. Phys. {\bf 17 }, 1468   \hfil\break

\nonfrenchspacing

\end